\documentclass[twocolumn,floatfix,prb,aps,showpacs]{revtex4-1}
\usepackage{graphicx,amsmath,amssymb,color}
\usepackage{nicefrac}
\usepackage[titletoc,title]{appendix}

\newcommand{\be}{\begin{equation}}
\newcommand{\ee}{\end{equation}}

\newcommand{\ba}{\begin{eqnarray}}
\newcommand{\ea}{\end{eqnarray}}

\begin{document}
\title{Parton construction of particle-hole-conjugate Read-Rezayi\\ parafermion fractional quantum Hall states and beyond}
\author{Ajit C. Balram$^{1}$, Maissam Barkeshli$^{2}$, and Mark S. Rudner$^{1}$}
\affiliation{$^{1}$Niels Bohr International Academy and the Center for Quantum Devices, Niels Bohr Institute, University of Copenhagen, 2100 Copenhagen, Denmark}
\affiliation{$^{2}$Condensed Matter Theory Center and Joint Quantum Institute, Department of Physics, University of Maryland, College Park, Maryland 20472 USA}
\date{\today}

\begin{abstract}
The Read-Rezayi (RR) parafermion states form a series of exotic non-Abelian fractional quantum Hall (FQH) states at filling $\nu = k/(k+2)$. Computationally, the wave functions of these states are prohibitively expensive to generate for large systems. We introduce a series of parton states, denoted ``$\bar{2}^{k}1^{k+1}$,'' and show that they lie in the same universality classes as the particle-hole-conjugate RR (``anti-RR'') states. Our analytical results imply that a $(U(1)_{k+1} \times U(2k)_{-1})/(SU(k)_{-2} \times U(1)_{-1})$ coset conformal field theory describes the edge excitations of the $\bar{2}^{k}1^{k+1}$ state, suggesting non-trivial dualities with respect to previously known descriptions. The parton construction allows wave functions in anti-RR phases to be generated for hundreds of particles. We further propose the parton sequence ``$\bar{n}\bar{2}^{2}1^{4}$,'' with $n=1,2,3$, to describe the FQH states observed at $\nu=2+1/2,~2+2/5$ and $2+3/8$.
\end{abstract}
\maketitle

The fractional quantum Hall effect (FQHE)~\cite{Tsui82} has revealed a variety of emergent many-body quantum phases that host exotic topological excitations. An important development in the field came about by a proposal of Moore and Read that the ``$5/2$'' state observed in the half-filled second Landau level (SLL) of GaAs~\cite{Willett87} could be described by a ``Pfaffian'' wave function~\cite{Moore91}. The excitations of the Pfaffian state are Majorana quasiparticles that feature non-Abelian braiding statistics~\cite{Read00}. Subsequently, Read and Rezayi (RR) proposed a class of FQH states hosting more general parafermionic excitations, including exotic ``Fibonacci'' anyons~\cite{Read99,Rezayi09}. Intriguingly, systems hosting such non-Abelian excitations may be utilized for fault-tolerant quantum computation~\cite{Kitaev03,Kitaev06,Nayak08,Freedman02}. 

Here we are motivated by the FQHEs observed in GaAs at filling factors $\nu=2+2/3$, $2+1/2$, $2+2/5$ and $2+3/8$ (see Refs.~\onlinecite{Willett87,Xia04,Pan08,Choi08,Kumar10,Zhang12}). Numerical studies have produced strong evidence that the first three members of this sequence may be described by the particle-hole conjugates of ``$k$-cluster'' RR wave functions~\cite{Read99,Rezayi09} (abbreviated as aRR$k$, where aRR stands for ``anti-Read-Rezayi''), with $k = 1$, $2$, and $3$, respectively~\cite{Ambrumenil88,Balram13b,Morf98,Scarola02,Pakrouski15,Rezayi17,Wojs09,Zhu15,Mong15,Pakrouski16}. In particular, the results in Refs.~\onlinecite{Wojs09,Zhu15,Mong15,Pakrouski16} indicate that the ground state of the experimentally observed~\cite{Xia04,Pan08,Choi08,Kumar10,Zhang12} FQHE at filling factor  $2+2/5$ is well-described by an aRR state that hosts Fibonacci anyons. This suggests that the $12/5$ FQHE may provide a solid state platform for universal fault-tolerant quantum computation.

The wave function of a $k$-cluster RR state is obtained by symmetrizing over partitions of $N$ particles into $k$ clusters, where each cluster forms a Laughlin state~\cite{Laughlin83}. (Here $N$ and $k$ are positive integers with $N$ divisible by $k$.) Importantly, the operation of symmetrization is computationally expensive, making it difficult to numerically evaluate these states and study their properties for large systems. The RR states can alternatively be obtained by exact diagonalization of model Hamiltonians~\cite{Read99} or using Jack polynomials~\cite{Bernevig08}, but these procedures are also limited to small sizes ($N\lesssim 30$). Thus there is great impetus to find more efficient representations of wave functions~\cite{Zaletel12,Lee15,Repellin15,Gonzalez16} in these exotic phases, to enable their further study.

In this work we introduce the ``$\bar{2}^{k}1^{k+1}$'' family of parton wave functions~\cite{Jain89b},  which for each $k = \{1, 2, 3, \ldots\}$ provides a state at filling factor $\nu = 2/(k + 2)$ within the same universality class as the aRR$k$ state. These parton wave functions can be evaluated for hundreds of particles, and thus provide means to numerically investigate the properties of parafermions in large systems. The $k=1$ and $k=2$ members of this parton family map onto states that were previously shown to lie in the same phases as the particle-hole conjugates of the 1/3 Laughlin state~\cite{Wu93,Balram16b}, and of the 1/2 Pfaffian state~\cite{Balram18} (i.e., the ``anti-Pfaffian'' state~\cite{Levin07,Lee07}), respectively. Below we give numerical evidence, based on wave function overlaps and entanglement spectra, that the $\bar{2}^{k}1^{k+1}$ state with $k=3$ is topologically equivalent to the aRR$3$ state. Using the effective field theory that arises from the parton mean-field ansatz, we compute several topological properties of $\bar{2}^{k}1^{k+1}$, including its chiral central charge, ground state degeneracy on the torus, and anyon content, and show that they match those of the aRR$k$ state.

\emph{Background.}--- Throughout this work we assume a single component system, and consider an ideal setting with zero width, no LL mixing, and zero disorder. The problem of interacting electrons confined to a given LL can be equivalently treated as a problem of electrons residing in the lowest Landau level (LLL), interacting via an effective interaction~\cite{Haldane83}. Thus we employ wave functions that reside in the LLL, keeping in mind that they can describe the FQHE in any LL (in particular, the SLL). 

The wave function of the $N$-particle, $k$-cluster RR state $\Psi^{\text{RR}k}_{k/(k+2)}$ at filling factor $\nu = k/(k + 2)$ is~\cite{Read99,Rezayi09,Cappelli01}:
\begin{eqnarray}
\Psi^{\text{RR}k}_{k/(k+2)}&=&\mathbb{S}\left[\prod_{i_{1}<j_{1}}(z_{i_{1}}-z_{j_{1}})^2\cdots \prod_{i_{k}<j_{k}}(z_{i_{k}}-z_{j_{k}})^2 \right] \nonumber \\
&&\times \prod_{i<j}(z_{i}-z_{j}) \exp\left[-\sum_{i}\frac{|z_{i}|^{2}}{4\ell^2}\right],
\label{eq_RRk}
\end{eqnarray}
where $z_{i}$, with $i = \{1, \ldots, N\}$, is the two-dimensional coordinate of the $i^{\rm th}$ electron, written as a complex number. (For ease of notation, below we suppress the ubiquitous Gaussian factors from all wave functions.) The $N$ particles are partitioned into $k$ internally correlated ``clusters'' of $N/k$ particles, with the product $\prod_{i_{l}<j_{l}}(z_{i_{l}}-z_{j_{l}})^2$ describing the correlations within a given cluster, $l$. The symbol $\mathbb{S}$ denotes symmetrization over all such partitions. The corresponding $k$-cluster {anti}-RR state, $\Psi^{\text{aRR}k}_{2/(k+2)}$, is described by the wave function:
\begin{equation}
 \Psi^{\text{aRR}k}_{2/(k+2)} = \mathcal{P}_{\rm ph} [\Psi^{\text{RR}k}_{k/(k+2)}],
 \label{eq_aRRk}
\end{equation}
where $\mathcal{P}_{\rm ph}$ denotes the operation of particle-hole conjugation. Due to particle-hole conjugation, $\Psi^{\text{aRR}k}_{2/(k+2)}$ occurs at filling factor $\nu = 1 - k/(k + 2) = 2/(k + 2)$.

For numerical work we employ the compact spherical geometry introduced by Haldane~\cite{Haldane83}. In this geometry $N$ electrons move on the surface of a sphere in the presence of a radial magnetic field $B$, the source of which is a Dirac monopole of strength $2Q$ sitting at the center of the sphere~\cite{footnote:sphere_disc}. The total magnetic flux through the sphere of radius $R$ is $4\pi R^{2}B=2Q(h/e)$. The radius of the sphere is thus related to the magnetic length, $\ell=\sqrt{\hbar /(eB)}$, via $R=\sqrt{Q}\ell$. Due to the spherical symmetry the total orbital angular momentum $L$ and its $z$-component $L_{z}$ are good quantum numbers in this geometry. 

Gapped quantum Hall ground states are rotationally invariant, i.e., they are uniform on the sphere and have $L_z = L = 0$. 
At a given filling factor $\nu$, one may find a variety of candidate ground states featuring distinct types of topological order~\cite{Wen92}. Each candidate ground state is realized at a specific value of the total magnetic flux through the sphere, $2Q =\nu^{-1}N-\mathcal{S}$, which is offset from its value in the plane, $N/\nu$, by a rational number $\mathcal{S}$ called the shift~\cite{Wen92}. If two states occur at different shifts, then they must describe different phases. Note that the converse, however, does not hold: topologically distinct states may occur with the same shift.

Before moving on to our parton ansatz, for reference we summarize some of the key properties of the RR$k$ and aRR$k$ states defined in Eqs.~(\ref{eq_RRk}) and (\ref{eq_aRRk}). The $k$-cluster RR state in Eq.~(\ref{eq_RRk}) occurs at monopole strength $2Q=[k/(k+2)]^{-1}N-3$, corresponding to the shift $\mathcal{S}^{\text{RR}k}=3$. The topological order of $\Psi^{\text{RR}k}_{k/(k+2)}$ is furthermore exhibited through the quantized thermal Hall conductance that it supports, $\kappa^{\text{RR}k}_{xy} = 3k/(k+2)$, in units of $[\pi^2 k_{\rm B}^2 /(3h)]T$, where $k_B$ is Boltzmann's constant and $T$ is the system's temperature~\cite{Read99,Bishara08}. In contrast, the aRR$k$ states in Eq.~(\ref{eq_aRRk}) are characterized by the flux-particle relation $2Q =\left[(k+2)/2\right]N-(1-k)$, corresponding to shift $\mathcal{S}^{\text{aRR}k}=1-k$. The thermal Hall conductance supported by $\Psi^{\text{aRR}k}_{2/(k+2)}$ is given by $\kappa_{xy}^{\text{aRR}k} = 1-\kappa^{\text{RR}k}_{xy}=-2(k-1)/(k+2)$, again in units of $[\pi^2 k_{\rm B}^2 /(3h)]T$~\cite{Read99,Bishara08}. 

\emph{Parton states.}---  
We now define a family of parton states, denoted $\bar{2}^{k}1^{k+1}$, each of which lies in the same universality class as the corresponding aRR$k$ state, $\Psi^{\text{aRR}k}_{2/(k+2)}$. The $\bar{2}^{k}1^{k+1}$ parton wave function, $\Psi^{\bar{2}^{k}1^{k+1}}_{2/(k+2)}$, is formed from a product of integer quantum Hall (IQH) states: 
\begin{equation}
\Psi^{\bar{2}^{k}1^{k+1}}_{2/(k+2)} = \mathcal{P}_{\rm LLL} [\Phi^{*}_{2}]^{k} \Phi^{k+1}_{1}\sim \frac{[\Psi^{\rm CF}_{2/3}]^{k}}{\Phi_{1}^{k-1}},
\label{eq_RRk_parton}
\end{equation}
where $\Phi_n$ is the $\nu = n$ IQH wave function of $N$ particles, and $\mathcal{P}_{\rm LLL}$ denotes projection into the lowest Landau level. Here $\Psi^{\rm CF}_{2/3} =\mathcal{P}_{\rm LLL} \Phi^{*}_{2}\Phi^{2}_{1}$ denotes the $\nu = 2/3$ composite fermion (CF) wave function~\cite{Jain89}. The $\sim$ sign indicates that (for $k > 1$) the rightmost expression in Eq.~(\ref{eq_RRk_parton}) differs from that in the middle in the details of how the projection to the LLL is carried out. We do not expect such details of the projection to change the topological properties of the state~\cite{Balram16b,Mishmash18}.

Crucially, the wave function given on the right hand side of Eq.~(\ref{eq_RRk_parton}) can be efficiently evaluated for large systems. This is so because the constituent CF wave function $\Psi^{\rm CF}_{2/3}$ can be evaluated for hundreds of electrons using the so-called Jain-Kamilla projection~\cite{Jain97,Jain97b}, details of which can be found in the literature~\cite{Moller05,Jain07,Davenport12,Balram15a}. 

When mapped to the spherical geometry, the states given in Eq.~(\ref{eq_RRk_parton}) occur at monopole strength $2Q =\left[(k+2)/2\right]N-(1-k)$, corresponding to filling factor $\nu=2/(k+2)$ and shift $\mathcal{S}^{\bar{2}^{k}1^{k+1}}=1-k$. These fillings and shifts precisely match those of the aRR$k$ states described by Eq.~(\ref{eq_aRRk}). 
This observation suggests that the wave functions given in Eq.~(\ref{eq_RRk_parton}) could lie in the same phases as the corresponding aRR states. For $k=1$, the $\bar{2}1^{2}$ state described by Eq.~(\ref{eq_RRk_parton}) is precisely the $2/3$ CF state (see above); this state is almost identical to the particle-hole conjugate of the $1/3$ Laughlin state~\cite{Wu93,Balram16b}. In Ref.~\onlinecite{Balram18} we studied the $\bar{2}^21^3$ state, and showed that it lies in the anti-Pfaffian~\cite{Levin07,Lee07} universality class. Below we discuss the case for arbitrary values of $k$. 

{\it Numerical results.}--- We first provide numerical evidence to show that $\Psi^{\bar{2}^{k}1^{k+1}}_{2/5}$ with $k = 3$ lies in the same phase as $\Psi^{{\rm aRR}3}_{2/5}$ given in Eq.~(\ref{eq_aRRk}). In Table~\ref{overlaps_n_1_LL_2_3_cube_over_1square_parton} we show overlaps of the parton wave function $\Psi^{\bar{2}^{3}1^{4}}_{2/5}$ with $\Psi^{{\rm aRR}3}_{2/5}$, as well as with the numerically-obtained exact ground state using the second Landau level Coulomb pseudopotentials, $\Psi^{{\rm SLL}}_{2/5}$. We find that the parton wave function has a good overlap with the corresponding anti-RR state. Furthermore, both $\Psi^{{\rm aRR}3}_{2/5}$ and $\Psi^{\bar{2}^{3}1^{4}}_{2/5}$ display decent overlap with the SLL Coulomb ground state, $\Psi^{{\rm SLL}}_{2/5}$. Similar to the aRR$3$ state~\cite{Read99,Rezayi09,Wojs09,Zhu15,Mong15,Pakrouski16}, the $k=3$ parton state of Eq.~(\ref{eq_RRk_parton}) can thus serve as a good candidate to describe the quantum Hall liquid occurring at $\nu=12/5$. 

\begin{table}
\centering
\begin{tabular}{|c|c|c|c|c|}
\hline
$N$ & $2Q$ & $|\langle \Psi^{{\rm SLL}}_{2/5}|\Psi^{{\rm aRR}3}_{2/5} \rangle|$ &  $|\langle \Psi^{\bar{2}^{3}1^{4}}_{2/5}|\Psi^{{\rm aRR}3}_{2/5} \rangle|$ & $|\langle \Psi^{{\rm SLL}}_{2/5}|\Psi^{\bar{2}^{3}1^{4}}_{2/5} \rangle|$  \\ \hline
4  & 12   		&0.9854    & 0.9173 &	0.8362	\\ \hline
6  & 17  		&0.9022    & 0.9107 &	0.6797	\\ \hline
8  & 22 	 	&0.9836    & 0.8821 &	0.8252	\\ \hline
\end{tabular}
\caption{\label{overlaps_n_1_LL_2_3_cube_over_1square_parton} 
Overlaps of $N$-particle FQH states on a sphere with magnetic flux $2Q$ corresponding to that of the particle-hole conjugate of the $k=3$ Read-Rezayi state (aRR$3$). We compare the wave functions of the parton state $\Psi^{\bar{2}^{3}1^{4}}_{2/5}$ [Eq.~(\ref{eq_RRk_parton})], the aRR$3$ state $\Psi^{{\rm aRR}3}_{2/5}$ [Eq.~(\ref{eq_aRRk})], and the ground state obtained by exact diagonalization using the SLL Coulomb pseudopotentials, $\Psi^{{\rm SLL}}_{2/5}$. The numbers for $|\langle \Psi^{{\rm SLL}}_{2/5}|\Psi^{{\rm aRR}3}_{2/5} \rangle|$ were previously given in Refs.~\onlinecite{Read99,Rezayi09,Pakrouski15,Zhu15,Kusmierz18}.}
\end{table}

\begin{figure}[t]
\begin{center}
\includegraphics[width=1.0\columnwidth]{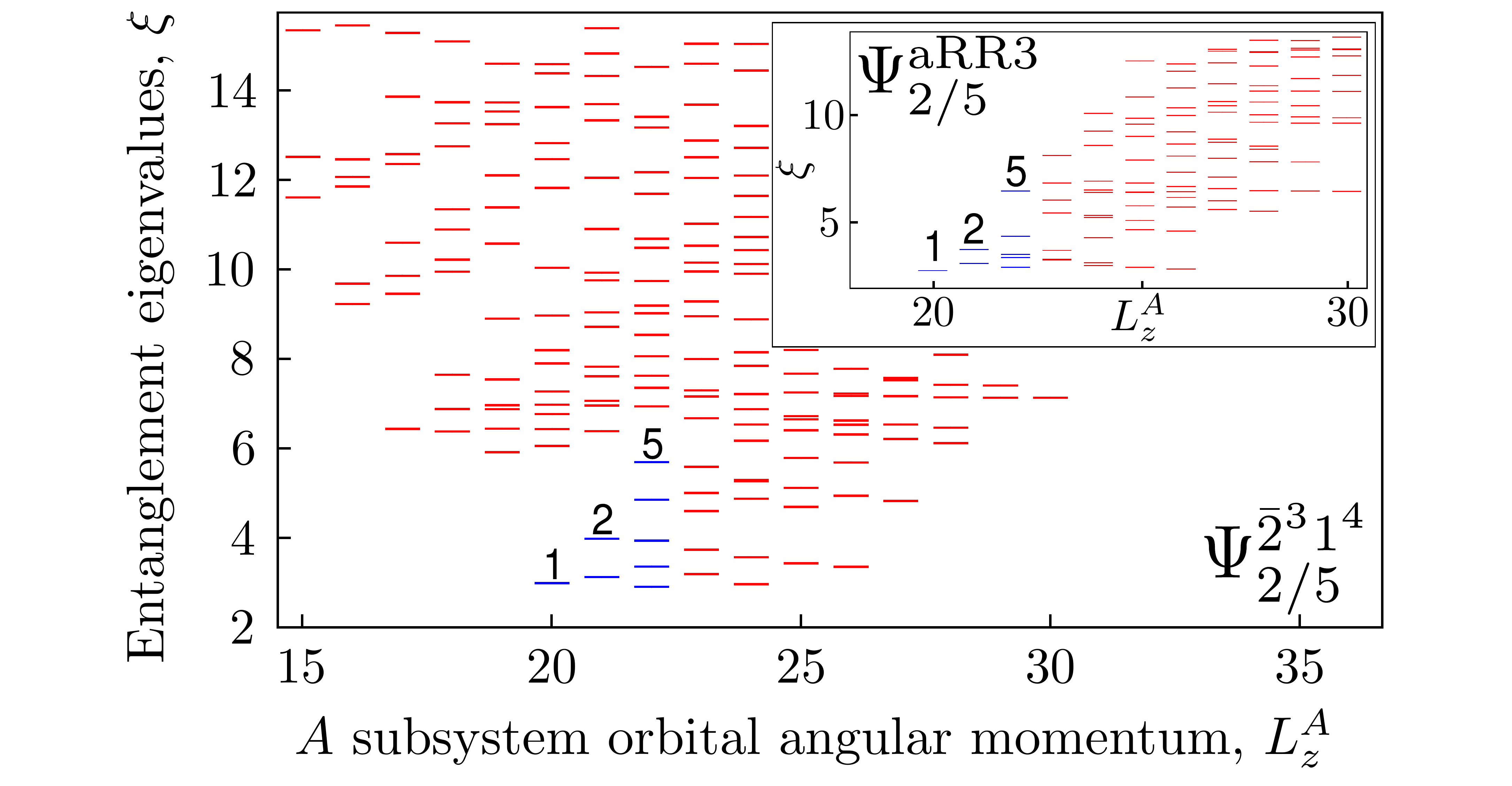}
\caption{(color online) Orbital entanglement spectrum of the $\Psi^{\bar{2}^{3}1^{4}}_{2/5}$ parton state for $N=8$ electrons at a flux $2Q=22$ on the sphere. The entanglement spectrum is calculated with respect to two subsystems, $A$ and $B$, with $N_{A}=N_{B}=4$ electrons and $l_{A}=12$ and $l_{B}=11$ orbitals, respectively. The entanglement levels are labeled by the $z$-component of the total orbital angular momentum of the $A$ subsystem, $L_{z}^{A}$. For comparison, in the inset we show the corresponding entanglement spectrum for $\Psi^{{\rm aRR}3}_{2/5}$. The multiplicities of low-lying levels (starting from $L_{z}^{A}=20$, going from left to right) are given by $1,2,5,\ldots$ and are identical for the two states.}
\label{fig:entanglement_spectra_parton_2_5}
\end{center}
\end{figure}

We provide further numerical evidence of the topological equivalence between $\Psi^{\bar{2}^{3}1^{4}}_{2/5}$ and $\Psi^{{\rm aRR}3}_{2/5}$ by comparing their entanglement spectra. The entanglement spectrum is a useful characterization tool, as it captures the structure of a FQH state's edge excitations~\cite{Li08}. The multiplicities of the low-lying entanglement levels carry a fingerprint of the topological order of the underlying state. Two states that lie in the same topological phase are expected to yield identical multiplicities. In Fig.~\ref{fig:entanglement_spectra_parton_2_5} we show the orbital entanglement spectrum~\cite{Haque07} of the $\bar{2}^{3}1^{4}$ state obtained on the sphere for a system of $N=8$ electrons at flux $2Q=22$. The multiplicities of the low-lying entanglement levels of $\Psi^{\bar{2}^{3}1^{4}}_{2/5}$ are identical to those of  $\Psi^{{\rm aRR}3}_{2/5}$. Thus we conclude that the $\bar{2}^{3}1^{4}$ parton state likely lies in the same phase as the aRR$3$ state. 

\emph{Field theory results.}---
Next, we consider the effective field theory that describes the associated parton mean-field ansatz (focusing on $k\geq 2$). Consider the following parton decomposition of the electron operator: $\wp = b f_1 \cdots f_k$, where the $f_i$ fields are fermions and $b$ is a boson (fermion) for $k$ odd (even). In the mean-field ansatz, $b$ forms a $\nu = 1/(k+1)$ Laughlin FQH state~\cite{Laughlin83}, while each fermion species $f_i$ forms a $\nu  = -2$ IQH state. This ansatz has a $U(1) \times SU(k)$ gauge symmetry. 

Integrating out the partons yields a non-Abelian Chern-Simons (CS) theory that we can use to explicitly compute the ground state degeneracy on the torus (see Supplemental Material (SM)~\cite{SM}). Carrying out this calculation for $k = 2,3,4,5,6,7$, we find a torus ground state degeneracy of $(k+1)(k+2)/2$, which agrees with the expected results for the aRR states. Using the field theory in combination with general consistency conditions from topological quantum field theory, we further demonstrate~\cite{SM} that the anyon content for the $k = 3$  parton state precisely matches that of aRR$3$. 
For $k > 3$ we derive a number of general properties for the anyon content of the parton states and show that they match with those of the corresponding aRR states~\cite{SM}.  

Finally, we consider the edge theory. The parton mean-field state (before implementing the gauge projection) is described by a $U(1)_{k+1} \times U(2k)_{-1}$ Wess-Zumino-Witten (WZW) conformal field theory (CFT) \cite{DiFrancesco97}. This CFT is comprised of $2k$ upstream-moving chiral fermion modes and $1$ downstream-moving chiral mode, giving a chiral central charge $c_{-,\rm MF} = -2k+1$. The gauge projection in the edge theory requires us to project out modes transforming non-trivially under the $U(1) \times SU(k)$ gauge symmetry, which leads to the $[U(1)_{k+1} \times U(2k)_{-1}]/[U(1)_{-1} \times SU(k)_{-2}]$ coset CFT \cite{DiFrancesco97,Wen99,SM}. The total central charge is $c_{-}=c_{-,\rm MF}-c_{-,\rm gauge}$, where $c_{-,\rm gauge} = -1-2(k^2-1)/(k+2)$ is the chiral central charge of the gauge degrees of freedom~\cite{SM}. We thus obtain $c_{-}=1-3k/(k+2)$, which precisely matches  the chiral central charge of the aRR$k$ state~\cite{Read99,Bishara08}.  

We note that a number of field theories for RR states have been described previously, such as an $SU(2)_k \times U(1)$ CS theory and a $U(1) \times Sp(k)_1$ CS theory~\cite{Barkeshli10,footnote:proper_definition_discrete_gauges}. The equivalence between these two theories is related to level-rank duality~\cite{DiFrancesco97, Barkeshli10}. Those results imply that the edge theory of the aRR states can be described by a $U(1)_1 \times SU(2)_{-k} \times U(1)$ WZW theory or, equivalently, by a dual $U(1)_1 \times Sp(k)_{-1} \times U(1)$ WZW theory. Our results imply that a $[U(1)_{k+1} \times U(2k)_{-1}]/[U(1)_{-1} \times SU(k)_{-2}]$ coset CFT can also describe the edge excitations of the aRR states, which suggests another non-trivial duality among these theories. We leave a detailed study of these dualities for future work. 

\begin{figure}[t]
\begin{center}
\includegraphics[width=1.0\columnwidth]{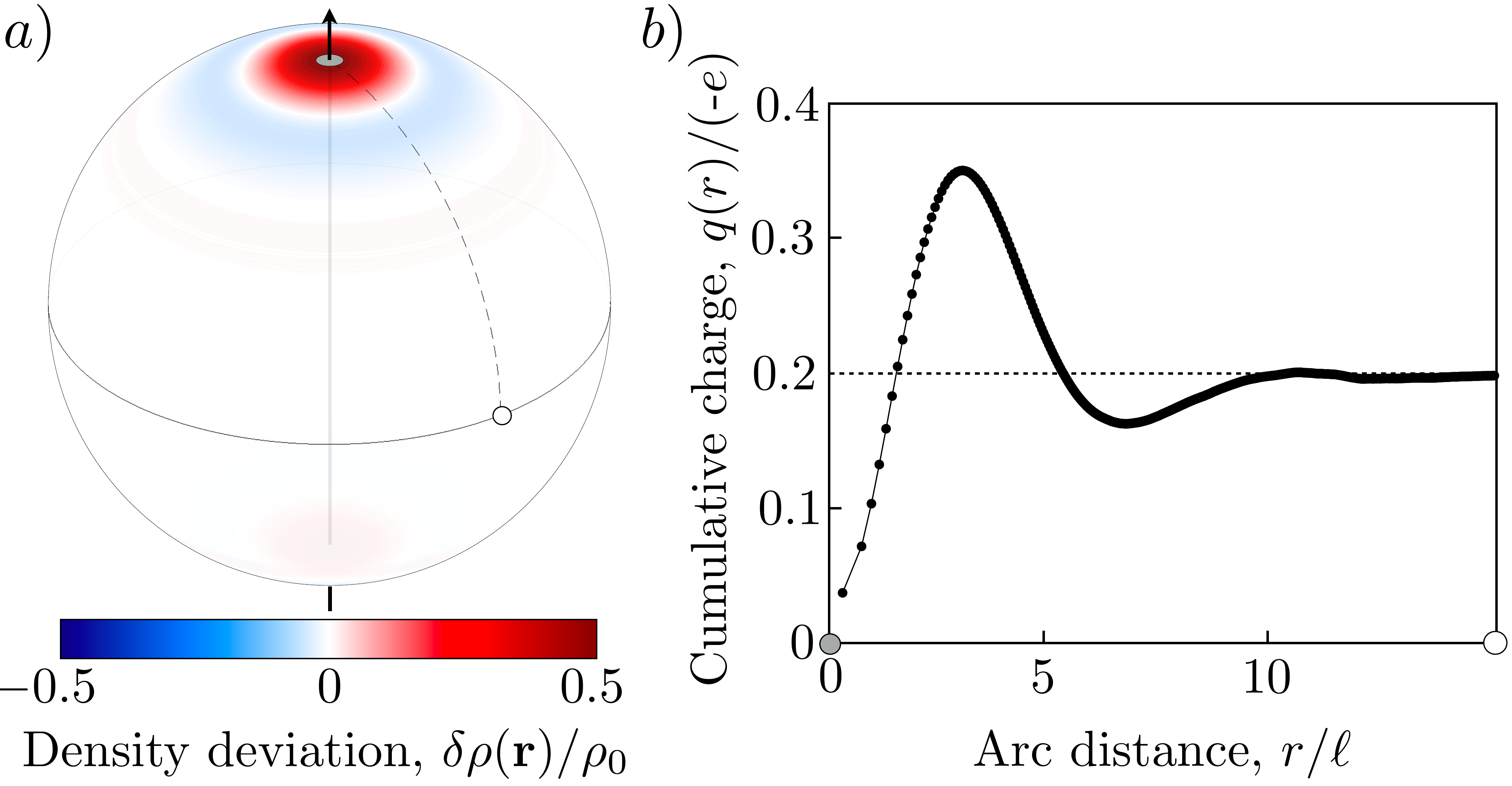}  
\caption{a) Density profile $\rho({\bf r})$ of a state with two far-separated quasiparticles at $\nu=2/5$, modeled by the parton wave function given in Eq.~(\ref{eq:wf_2fQPs_2_5}) for $N=80$ electrons on the sphere. The two quasiparticles are located at the north and south pole of the sphere. The color represents the density deviation from its value $\rho_0$ in the uniform $\Psi^{\bar{2}^{3}1^{4}}_{2/5}$ state: $\delta\rho({\bf r})/\rho_{0}=[\rho({\bf r})-\rho_{0}]/\rho_{0}$. b) The integrated cumulative charge $q(r)$ [see text for definition] as a function of latitude, parametrized by the distance $r$ along the arc from the north pole to the equator (in units of magnetic length, $\ell$).
The cumulative charge approaches the value $-0.2e$ near the equator.}
\label{fig:density_2fQPs_2bar2bar2bar1111}
\end{center}
\end{figure}

\emph{Discussion.}--- A major advantage of our parton wave functions is that they can be constructed for large systems. As a proof of principle, we numerically demonstrate that the smallest charge quasiparticle (QP) of the $2/5$ parton state carries a charge $-e/5$, where $-e < 0$ is the charge of the electron~\cite{Jain89b}. To this end, we create a model state at filling factor $\nu = 2/5$ with two far-separated QPs, one located at each pole of the sphere~\cite{footnote:two_QPs}:
\begin{equation}
 \Psi^{2\text{-}{\rm QPs}}_{2/5} = \mathcal{P}_{\rm LLL}[\Phi^{(2\text{-}{\rm h})}_{2}]^{*}[\Phi^{2}_{2}]^{*}\Phi^{4}_{1}
  \sim \frac{\Psi^{{\rm CF,}(2\text{-}{\rm QP})}_{2/3}[\Psi^{\rm CF}_{2/3}]^{2}}{\Phi_{1}^{2}},
  \label{eq:wf_2fQPs_2_5}
\end{equation}
where $\Phi^{(2\text{-}{\rm h})}_{2}$ and $\Psi^{{\rm CF,}(2\text{-}{\rm QP})}_{2/3}$ are the $\nu=2$ IQH and $\nu=2/3$ CF states with two holes or two QPs, respectively, located at opposite poles of the sphere. The CF wave functions are evaluated using the Jain-Kamilla method~\cite{Jain97,Jain97b,Moller05,Jain07,Davenport12,Balram15a}. In Fig.~\ref{fig:density_2fQPs_2bar2bar2bar1111}a we show the density profile $\rho({\bf r})$ of $\Psi^{2\text{-}{\rm QPs}}_{2/5}$ for $N = 80$ electrons. Close to the equator, the density approaches the value $\rho_{0}$ of the uniform $\bar{2}^{3}1^{4}$ state. To extract the QP charge, we integrate the deviation of the charge density from its uniform value, $\rho({\bf r})-\rho_{0} \equiv  \delta\rho({\bf r})$, over the northern hemisphere. In Fig.~\ref{fig:density_2fQPs_2bar2bar2bar1111}b we plot the cumulative charge $q(r) = \int_0^r \delta \rho({\bf r'})d^{2}{\bf r'}$ as a function latitude, parametrized by the arc distance $r$ along the dashed contour shown in Fig.~\ref{fig:density_2fQPs_2bar2bar2bar1111}a. From the limiting value of $q(r)$ at the equator we extract a charge of $-0.197e$, which is close to the expected value of $-0.2e$ (attained in the thermodynamic limit when the QPs do not overlap). 

Building on our results for the $k=3$ case, we are led to consider a new ``$\bar{n}\bar{2}^{2}1^{4}$'' parton sequence described by the wave functions:
\begin{equation}
\Psi^{\bar{n}\bar{2}^{2}1^{4}}_{n/(3n-1)} = \mathcal{P}_{\rm LLL} [\Phi^{*}_{n}]  [\Phi^{*}_{2}]^{2} \Phi^{4}_{1}\sim \frac{\Psi^{\rm CF}_{n/(2n-1)}[\Psi^{\rm CF}_{2/3}]^{2}}{\Phi^{2}_{1}}.
\label{eq:parton}
\end{equation}
In the spherical geometry, $\Psi^{\bar{n}\bar{2}^{2}1^{4}}_{n/(3n-1)}$ occurs at monopole strength $2Q =\left[(3n-1)/n\right]N+n$ and hence has filling factor $\nu=n/(3n-1)$ and shift $\mathcal{S}^{\bar{n}\bar{2}^{2}1^{4}}=-n$. We thus obtain states at filling factors $\nu=1/2,\, 2/5,\, 3/8, \ldots$ for $n=1,2,3,\ldots$, respectively. The $n=1$ member of this sequence likely lies in the same universality class as the $\bar{2}^{2}1^{3}$ state~\cite{footnote:mod_phi1_square_factors}, which we showed in a previous work lies in the anti-Pfaffian phase~\cite{Balram18}. We discussed the $n=2$ case in detail in this paper and concluded that it lies in the same phase as the aRR3 state. Intriguingly, the $n=3$ state of Eq.~(\ref{eq:parton}) provides a candidate ground state wave function that could possibly describe the FQHE at $2+3/8$~\cite{Xia04,Pan08,Choi08,Kumar10,Zhang12,Toke08,Hutasoit16}. We thus speculate that the $\bar{n}\bar{2}^{2}1^{4}$ family of parton states may capture the observed plateaus at $2/5$ and $3/8$ in the SLL of GaAs~\cite{Xia04,Pan08,Choi08,Kumar10,Zhang12} that were not covered in the $\bar{n}\bar{2}1^{3}$ sequence of Ref.~\onlinecite{Balram18a}. Generically, we find that the parton states of Eq.~(\ref{eq:parton}) are topologically different from other families of candidate states occuring at the same $n/(3n-1)$ sequence of filling factors~\cite{SM}.

Although we only considered states with a single component, our parton construction can be extended in a straightforward manner to build multi-component states at the corresponding filling factors, where the different components could represent either the spin, valley or orbital degrees of freedom. The properties of these states remain to be explored. 

Taken together with our previous works~\cite{Balram18,Balram18a}, the results presented in this article suggest that almost all fractional quantum Hall states observed in the second Landau level of GaAs could be described by the $\bar{n}\bar{2}1^{3}$ or $\bar{n}\bar{2}^{2}1^{4}$ parton ansatz, with $n=1,2,3$, or their particle-hole conjugates. The states observed in the second LL that do not fall in these sequences or their particle-hole conjugates, e.g., at filling factor 1/5 and 2/7, are likely well described by composite fermion states~\cite{Ambrumenil88,Kusmierz18} (which are also parton states). In all, except for the lowest Landau level states at $\nu=4/11$ and $5/13$~(see, e.g., Refs.~\onlinecite{Pan03,Pan15,Samkharadze15b}), it appears that all fractional quantum Hall states observed to date (or their particle-hole conjugates), including in graphene~\cite{Xu09,Bolotin09,Feldman13,Amet15,Balram15c,Zeng18,Kim18} and wide quantum wells~\cite{Luhman08,Shabani09a,Shabani09b,Shabani13,Faugno19}, admit simple parton descriptions. 

\begin{acknowledgments}
The Center for Quantum Devices is funded by the Danish National Research Foundation. This work was supported by the European Research Council (ERC) under the European Union Horizon 2020 Research and Innovation Programme, Grant Agreement No. 678862. A.C.B. and M.R. also thank the Villum Foundation for support. MB is supported by NSF CAREER (DMR-1753240), JQI-PFC-UMD and an Alfred P. Sloan Research Fellowship. Some of the numerical calculations were performed using the DiagHam package, for which we are grateful to its authors. 
\end{acknowledgments}

\bibliography{biblio_fqhe}

\begin{thebibliography}{86}
\bibitem[{foo({\natexlab{a}})}]{footnote:sphere_disc}
\bibinfo{note}{We follow the convention of writing wave functions in the planar
  disc geometry, as in Eqs.~(\ref{eq_RRk}) and (\ref{eq_aRRk}), which can be
  adapted to the spherical geometry using the stereographic projection.}
\bibitem[{foo({\natexlab{b}})}]{footnote:proper_definition_discrete_gauges}
\bibinfo{note}{Defining these theories properly requires incorporating some
  discrete gauge symmetries as well, which we do not describe here.}
\bibitem[{foo({\natexlab{c}})}]{footnote:two_QPs}
\bibinfo{note}{The state $\Phi_{2}$ can only be constructed for an even number
  of particles on the sphere, thus precluding the construction of a state with
  a single QP at $\nu=2/5$ using our parton ansatz.}
\bibitem[{foo({\natexlab{d}})}]{footnote:mod_phi1_square_factors}
\bibinfo{note}{The $\bar{1}\bar{2}^{2}1^{4}$ and $\bar{2}^{2}1^{3}$ parton
  states differ from each other only by a factor of $|\Phi_{1}|^{2}$. Such
  factors of $|\Phi_{1}|^{2}$ are expected to only weakly affect the resulting
  LLL-projected wave function~\cite{Balram16b,Mishmash18}.}
\end{thebibliography}
\bibliographystyle{apsrev_nourl}

\newpage
\null\thispagestyle{empty}
\newpage

\widetext
\begin{center}
\textbf{\large Supplemental Material for ``Parton construction of particle-hole-conjugate Read-Rezayi parafermion fractional quantum Hall states and beyond''} 
\end{center}

\setcounter{figure}{0}
\setcounter{table}{0}
\setcounter{equation}{0}
\renewcommand\thefigure{S\arabic{figure}}
\renewcommand\thetable{S\arabic{table}}
\renewcommand\theequation{S\arabic{equation}}

In this Supplemental Material (SM), we discuss in detail the effective field theory of the $\bar{2}^{k}1^{k+1}$ parton states considered in the main text. We show that the the $\bar{2}^{k}1^{k+1}$ parton state is topologically equivalent to the particle-hole conjugate of the $k$-cluster Read-Rezayi (RR$k$) state, referred to as the anti-RR$k$ (aRR$k$) state. In particular, we show that ground state degeneracy on a torus, the shift, the chiral central charge, the charges of the various quasiparticles, and their fusion rules, are identical for the two states. For the $k = 3$ case we demonstrate that all quasiparticles match exactly. \\

In Sec. \ref{sec:comparison_parton_others} we compare the ``$\bar{n}\bar{2}^{2}1^{4}$'' set of parton states proposed in the main text with other known families of candidate states occurring at the same sequence $\nu=n/(3n-1)$ of filling factors. We find that, generically, the members of the $\bar{n}\bar{2}^{2}1^{4}$ parton sequence are topologically distinct from the members of the other known families of candidate states proposed at the same filling factor.

\section{Particle-hole conjugates of Read-Rezayi states~\cite{Read99}}
In this section, we discuss key properties of the aRR states.
The RR states are a series of non-Abelian fractional quantum Hall (FQH) states at filling fractions $\nu = k/(k+2)$. 
Their particle-hole conjugates, which we refer to as the anti-RR states, occur at filling fractions 
\begin{align}
\nu = 1 - \frac{k}{k+2} = \frac{2}{k+2}. 
\end{align}

The chiral central charge of the RR states is $c_{-}= 3k/(k+2)$~\cite{Read99,Bishara08}. In contrast, the chiral central charge of the anti-RR states is 
\begin{align}
c_{-} = 1 - \frac{3k}{k+2}=-\frac{2(k-1)}{k+2}. 
\end{align}
From the point of view of topological order (i.e. fusion and braiding of the quasiparticles), the RR and anti-RR states are chirality-reversed counterparts of each other. 
They exhibit the same numbers of quasiparticles, with identical fusion rules, and their topological twists are complex conjugates of each other. In particular, the ground state degeneracy on a torus is~\cite{Read99}:
\begin{align*}
\text{Torus ground state degeneracy}=\frac{1}{2} (k+1)(k+2). 
\end{align*}
The non-Abelian fusion rules of the quasiparticles are governed by the fusion rules of $SU(2)_k$ Chern-Simons (CS) theory. 

\subsection{Quasiparticle structure for the $k = 3$ RR state}
\label{3RRsec}
In the case $k  =3$, the Read-Rezayi states have $N_{\rm qp}=10$ quasiparticles. In the edge theory, the electron operator can be written as
\begin{align}
\Psi_e = \psi e^{i \sqrt{\nu^{-1}} \varphi} = \psi e^{i \sqrt{(k+2)/k} \varphi} = \psi e^{i \sqrt{5/3} \varphi} ,
\end{align}
where $\varphi$ is a chiral boson and $\psi$ is the simple current of the $\mathbb{Z}_3$ parafermion conformal field theory (CFT), which has fractional scaling dimension $h_\psi = (k-1)/k = 2/3$ and satisfies $\psi^{k}=\psi^{3}=1\implies \Psi^{3}_e=e^{i 3\sqrt{5/3} \varphi}$~\cite{Zamolodchikov85}. The quasiparticle operators in the edge theory consist of the operators in the $\mathbb{Z}_3$ parafermion $\times $ $U(1)$ CFT that are local with respect to the electron operator. 
This gives five \emph{topologically distinct} Abelian particles described by fields $V^{{\rm RR}3}_a$, for $a=0,1,2,3,4$, with:
\begin{align}
&V^{{\rm RR}3}_a = e^{i a 3/5 \sqrt{5/3} \varphi},  \;\; Q^{{\rm RR}3}_a = 3a/5, \\
&\;\; \theta^{{\rm RR}3}_a = 2\pi h^{{\rm RR}3}_a = \pi a^2 3/5.
\end{align}
Here $Q_a$ is the fractional charge and $\theta_a$ describes the exchange statistics (i.e.,~the topological spin). 
We can see that $V^{{\rm RR}3}_5$ is trivial because it corresponds to $\Psi_e^3$. These values are tabulated in Table~\ref{3RRtable}. 

The non-Abelian particles are described by:
\begin{align}
\tau V^{{\rm RR}3}_a,Q^{{\rm RR}3}_{\tau V^{{\rm RR}3}_a} = 3a/5,~\theta^{{\rm RR}3}_{\tau V^{{\rm RR}3}_a} = \theta^{{\rm RR}3}_a +  \theta^{{\rm RR}3}_\tau,
\end{align}
where $\tau \times \tau = 1+ \tau$ is a Fibonacci particle, with $\theta^{{\rm RR}3}_{\tau} = 2\pi (2/5)$. 
\begin{table}
\centering
\begin{tabular}{l | r | r r}
\hline
 Label & Charge, $Q$ (mod 2)& Twist, $e^{i \theta}$ \\
\hline
$V^{{\rm RR}3}_0$  & $0$ & $1$ &      \\
$V^{{\rm RR}3}_1$  & $3/5$ & $e^{i \pi 3/5}$ &      \\
$V^{{\rm RR}3}_2$  & $1 + 1/5$ & $e^{i \pi 2/5}$ &      \\
$V^{{\rm RR}3}_3$  & $1+ 4/5$ & $-e^{i \pi 2/5}$ &      \\
$V^{{\rm RR}3}_4$  & $2/5$ & $-e^{i \pi 3/5}$ &      \\
$V^{{\rm RR}3}_5 \sim \Psi_e$ & $1$ & $-1$ & \\
$V^{{\rm RR}3}_6 \sim V^{{\rm RR}3}_1 \Psi_e$ & $1 + 3/5$ & $-e^{i\pi 3/5}$ & \\
$V^{{\rm RR}3}_7 \sim V^{{\rm RR}3}_2 \Psi_e$ & $1/5$ & $-e^{i\pi 2/5}$ & \\
$V^{{\rm RR}3}_8 \sim V^{{\rm RR}3}_3 \Psi_e$ & $4/5$ & $e^{i\pi 2/5}$ & \\
$V^{{\rm RR}3}_9 \sim V^{{\rm RR}3}_4 \Psi_e$ & $1+2/5$ & $e^{i \pi 3/5}$ & \\
\hline
\end{tabular}
\caption{Abelian particles of $k = 3$ RR state. Charge $2$ bosons are treated as completely trivial. Charge $1$ fermion (electron) is kept track of. \label{3RRtable} }
\end{table}

\subsection{Quasiparticle structure for $k = 3$ anti-Read-Rezayi state}
\label{3aRRsec}
The particle-hole conjugate of the $k = 3$ RR state has Abelian quasiparticles described by:
\begin{align}
&V^{{\rm aRR}3}_a = e^{i a 3/5 \sqrt{5/3} \varphi},  \;\; Q^{{\rm aRR}3}_a = 3a/5, \;\; \\
&\theta^{{\rm aRR}3}_a = 2\pi h^{{\rm aRR}3}_a = -\pi a^2 3/5 . 
\nonumber 
\end{align}
These values are tabulated in Table~\ref{3antiRRtable}. \\
The non-Abelian particles are described by: 
\begin{align}
&\tau V^{{\rm aRR}3}_a, \;\; Q^{{\rm aRR}3}_a = 3a/5, \;\; \\
& \theta^{{\rm aRR}3}_{\tau V^{{\rm aRR}3}_a} = \theta^{{\rm aRR}3}_a +  \theta^{{\rm aRR}3}_\tau = -\pi a^2 3/5 - 2\pi 2/5,
\end{align}
where now $\theta^{{\rm aRR}3}_\tau = -2\pi (2/5)$.

\begin{table}[t]
\centering
\begin{tabular}{l | r |r r  }
\hline
Label & Charge, $Q$ (mod 2) & Twist, $e^{i \theta}$ \\
\hline
$V^{{\rm aRR}3}_0$  & $0$ & $1$ &      \\
$V^{{\rm aRR}3}_1$  & $3/5$ & $-e^{i \pi 2/5}$ &      \\
$V^{{\rm aRR}3}_2$  & $1 + 1/5$ & $-e^{i \pi 3/5}$ &      \\
$V^{{\rm aRR}3}_3$  & $1+ 4/5$ & $e^{i \pi 3/5}$ &      \\
$V^{{\rm aRR}3}_4$  & $2/5$ & $e^{i \pi 2/5}$ &      \\
$V^{{\rm aRR}3}_5 \sim \Psi_e$ & $1$ & $-1$ & \\
$V^{{\rm aRR}3}_6 \sim V^{{\rm aRR}3}_1 \Psi_e$ & $1 + 3/5$ & $e^{i\pi 2/5}$ & \\
$V^{{\rm aRR}3}_7 \sim V^{{\rm aRR}3}_2 \Psi_e$ & $1/5$ & $e^{i\pi 3/5}$ & \\
$V^{{\rm aRR}3}_8 \sim V^{{\rm aRR}3}_3 \Psi_e$ & $4/5$ & $-e^{i\pi 3/5}$ & \\
$V^{{\rm aRR}3}_9 \sim V^{{\rm aRR}3}_4 \Psi_e$ & $1+2/5$ & $-e^{i \pi 2/5}$ & \\
\hline
\end{tabular}
\caption{Abelian particles of $k = 3$ anti-RR state. Charge $2$ bosons are treated as completely trivial. Charge $1$ fermion (electron)
is kept track of. \label{3antiRRtable} }
\end{table}

\section{Properties of the $\Phi_{\bar 2}^k \Phi_1^{k+1}$ FQH states}
Consider the following non-projected version of the family of wave functions discussed in the main text:
\begin{align}
\Psi^{{\bar 2}^k 1^{k+1}} = \Phi_{\bar 2}^k \Phi_1^{k+1},
\label{wfnAnsatz}
\end{align}
where $\Phi_n$ is a $\nu = n$ integer quantum Hall (IQH) wave function, while $\Phi_{\bar n} = \Phi_n^*$ is a $\nu = -n$ IQH wave function. Our goal is to determine the topological order associated with this family of wave functions. In particular, we wish to show that these wave functions correspond to the particle-hole conjugates of the $k$-cluster RR states~\cite{Read99}. 

\subsection{Parton construction}

The state in Eq.~(\ref{wfnAnsatz}) can be obtained from a parton construction
\begin{align}
\wp = f_1 \cdots f_k~\psi_1 \cdots \psi_{k+1},
\end{align}
where $f_1, \cdots, f_k$ form $\nu = -2$ IQH states, while $\psi_1,\cdots,\psi_{k+1}$ form $\nu = 1$ IQH states. This construction has a gauge
group $SU(k) \times SU(k+1) \times U(1)$. 

Alternatively, we can consider the parton construction
\begin{align}
\wp = f_1 \cdots f_k~b,
\end{align}
where $f_i$ form $\nu  = -2$ IQH states, while $b$ forms a $\nu = 1/(k+1)$ Laughlin state~\cite{Laughlin83}. Note that for $k$ odd, $b$ is a boson, while for $k$ even, it is a fermion. This ansatz has a $SU(k)\times U(1)$ gauge symmetry. 

\subsection{Effective field theory}
Without loss of generality, we choose the charges of the parton fields under the background $U(1)$ electromagnetic gauge field $A$ such that $b$ carries charge $1$, while all other fields carry charge $0$. 
Let $a$ denote the $U(1)$ gauge field under which $b$ and $f_1$ carry charge $1$ and $-1$, respectively.
The remaining gauge field is an $SU(k)$ gauge field $A_{SU(k)}$, which is an element of the Lie algebra of $SU(k)$: 
\begin{align}
A_{SU(k)} = A_{SU(k)}^\alpha T^\alpha, \;\;\; \alpha = 1, \cdots, k^2 - 1.
\end{align}
Here $T^\alpha$ is a $k \times k$ matrix in the fundamental representation of the $SU(k)$ Lie algebra.

The effective Lagrangian of the theory can therefore be written as
\begin{align}
\mathcal{L} = -\frac{k+1}{4\pi} \tilde{a} \partial \tilde{a} + \frac{1}{2\pi}(a + A)\partial \tilde{a} + \mathcal{L}(f_1,\cdots, f_k, A_{SU(k)}, a),
\end{align}
where $\tilde{a}$ is the gauge field that describes the $\nu=1/(k+1)$ Laughlin state of the parton $b$. Since $a$ couples only to $f_1$, we can define another generator
\begin{align}
T^0_{ij} = \delta_{i1}\delta_{j1},
\end{align}
so that we can couple the fermion vector $f^\dagger = (f_1^\dagger, \cdots, f_k^\dagger)$ to the matrix valued gauge field
$a T^0$. Therefore the fermions $f$ couple to the gauge field
\begin{align}
\mathcal{A} = \sum_{i=0}^{k^{2}-1}\mathcal{A}^i T^i = a T^0 + \sum_{\alpha = 1}^{k^2-1} A_{SU(k)}^\alpha T^\alpha,
\end{align} 
where we define $\mathcal{A}^0 \equiv a$.

Integrating out the fermions $\{f_1, \ldots, f_k\}$, each of which forms a $\nu = -2$ IQH state, gives the effective action:
\begin{align}
\mathcal{L} = -\frac{k+1}{4\pi} \tilde{a} \partial \tilde{a} + \frac{1}{2\pi}(a + A)\partial \tilde{a} - 2 \mathcal{L}_{CS}(\mathcal{A}),
\end{align}
where $\mathcal{L}_{CS}(\mathcal{A})$ is the Chern-Simons Lagrangian for $\mathcal{A}$: 
\begin{align}
\mathcal{L}_{CS}(\mathcal{A}) = \frac{1}{4\pi} \epsilon^{\mu\nu\lambda} [\mathcal{A}_\mu^a \partial_\nu \mathcal{A}_\lambda^b \text{ Tr } T^a T^b
+ \frac{2}{3} \mathcal{A}_\mu^a \mathcal{A}_\nu^b \mathcal{A}_\lambda^c \text{ Tr } T^a T^b T^c].  
\end{align} 
 This (2+1)D bulk effective action can be used to study a number of properties of the state (\ref{wfnAnsatz}), such as its filling fraction and the ground state degeneracy on a torus.

\subsubsection{Ground state degeneracy on a torus}
In order to compute the ground state degeneracy on a torus, we first note that the CS theory imposes the constraint that the gauge fields must all be flat connections on the torus (i.e., the magnetic field piercing the torus is identically zero everywhere). Non-trivial gauge configurations are therefore specified by the Wilson loops $W_\alpha$ and $W_\beta$ along the two independent non-contractible cycles of the torus, $\alpha$ and $\beta$. 
Mathematically, inequivalent flat connections on a torus are specified by maps $(\text{Hom}: \pi_1(T^2) \rightarrow G)/G$, which are homomorphisms from the fundamental group of the torus into $G$, modulo $G$. Here $G = U(1) \times SU(k) \times U(1)$ is the gauge group of the Chern-Simons theory. Since $\pi_1(T^2)$ is Abelian, we can perform simultaneous gauge transformations so that $W_\alpha$ and $W_\beta$ lie in the maximal Abelian subgroup, $G_{abl}$ of the gauge group $G$. $G_{abl}$ is usually referred to as the maximal torus of $G$. The maximal torus is generated by the Cartan subalgebra of the Lie algebra of $G$. 

In order to restrict the gauge fields to lie in the Cartan subalgebra of the Lie algebra of the gauge group, we take
\begin{align}
\mathcal{A} = \mathcal{A}^I p^I,
\end{align}
where $\{p^I\}$ for $I = 0,\cdots, k-1$ are $k \times k$ diagonal matrices in the Cartan subalgebra of $SU(k) \times U(1)$. 
These are:
\begin{align}
p^0 &= {\rm diag}(1,0,\cdots, 0),
\nonumber \\
p^1 &= {\rm diag}(1,-1,\cdots, 0),
\nonumber \\
p^2 &= {\rm diag}(0,1,-1,\cdots,0),
\nonumber \\
p^{k-1} &= {\rm diag}(0,\cdots,0,1,-1). 
\end{align}

When restricted to the Cartan subalgebra, the effective Lagrangian becomes
\begin{align}
\mathcal{L} &= -\frac{k+1}{4\pi} \tilde{a} \partial \tilde{a} + \frac{1}{2\pi}(a + A)\partial \tilde{a} - 2 \frac{1}{4\pi} \mathcal{A}^I \partial \mathcal{A}^J Tr p^I p^J 
\nonumber \\ 
&= - \frac{1}{4\pi}\tilde{K}_{IJ} \tilde{\mathcal{A}}^I \partial \tilde{\mathcal{A}}^J + \frac{1}{2\pi} q^I A \partial \tilde{\mathcal{A}}^I . 
\end{align}
Here $\tilde{\mathcal{A}} = (\tilde{a}, a, \mathcal{A}^1, \cdots, \mathcal{A}^{k-1})$. As such, $\tilde{K}$ is a $(k+1) \times (k+1)$ dimensional matrix, such that 
\begin{align}
\tilde{K}_{11} &= k+1
\nonumber \\
\tilde{K}_{22} &= 2
\nonumber \\
\tilde{K}_{12} = \tilde{K}_{21} &= -1
\nonumber \\
\tilde{K}_{23} = \tilde{K}_{32} &= 2,
\end{align}
and
\begin{align}
\tilde{K}_{IJ} = 2 K_{IJ} , \;\;  3 \leq I,J \leq k+1,
\end{align}
where $K$ is the Cartan matrix of $SU(k)$, which is a $(k - 1) \times (k-1)$ dimensional matrix:
\begin{align}
K_{IJ} = 2 \delta_{IJ} - \delta_{I, J+1} - \delta_{I,J-1} . 
\end{align}
The charge vector is given by $\vec{q} = (1, 0,0, \cdots, 0)$. 

The theory has large gauge transformations
\begin{align}
f \rightarrow U_I f,
\end{align}
with 
\begin{align}
U_I = e^{2\pi i p^I x_i/L},
\end{align}
where $x_1$ and $x_2$ are the coordinates in the two directions of the torus, and $L$ is the length of each side. 
With the above normalization of the $\{p^I\}$, these are the minimal large gauge transformations. 

In addition to the large gauge transformations, there are discrete gauge transformations $W \in SU(k)$
which keep the Abelian subgroup unchanged but interchange the $\mathcal{A}^I$ amongst themselves. These satisfy
\begin{align}
W^\dagger G_{abl} W = G_{abl},
\end{align}
or, alternatively, allow us us define 
\begin{align}
W^\dagger p^I W = \tilde{W}_{IJ} p^I,
\end{align}
for some $k \times k$ matrix $\tilde{W}$. These discrete transformations originate from the $k!$ permutations of the fermions $f$. 

With the above gauge transformations in mind, we now quantize the theory. We pick the gauge 
$\tilde{A}^I_0 = 0$. The ground state degeneracy is determined by the zero modes of the gauge field, which we thus parametrize as:
\begin{align}
\mathcal{A}_i^I = \frac{2\pi}{L} X_i^I .
\end{align}
The Lagrangian becomes
\begin{align}
\mathcal{L} = 2\pi \tilde{K}_{IJ} X_1^I \dot{X}_2^J.
\end{align}
The Hamiltonian vanishes. The conjugate momentum to $X_2^J$ is
\begin{align}
p_2^J = 2\pi \tilde{K}_{IJ} X_1^I. 
\end{align}
Since $X_2^J \sim X_2^J + 1$ as a result of the large gauge transformations, we can write the wave functions
as
\begin{align}
\psi(\vec{X}_2) = \sum_{\vec n} c_{\vec n} e^{i 2 \pi \vec{n} \cdot \vec{X}_2},
\end{align}
where $\vec{X}_2 = (X_2^1, \cdots, X_2^{k+1})$, and $\vec{n}$ is a $(k+1)$-dimensional vector of integers. In momentum space the wave function is 
\begin{align}
\phi(\vec{p}_2) &= \sum_{\vec n} c_{\vec n} \delta^{(k+1)}(\vec{p}_2 - 2\pi \vec{n})
\nonumber \\
&\sim \sum_{\vec{n}} c_{\vec n} \delta^{(k+1)}(K \vec{X}_1 - \vec{n}),
\end{align}
where $\delta^{(k+1)}(\vec{x})$ is a $(k+1)$-dimensional delta function. Since $X_1^J \sim X_1^J + 1$,
it follows that $c_{\vec n} = c_{\vec n'}$, where $(\vec{n}')^I = n^I + K_{IJ}$, for any $J$. Furthermore, each
discrete gauge transformation $W_i$ that keeps the Abelian subgroup $G_{abl}$ invariant corresponds to a matrix
$\tilde{W}_i$, which acts on the diagonal generators. These imply the equivalences $c_{\vec n} = c_{\tilde{W} \vec{n}}$.
The number of independent $c_{\vec n}$ can now be computed for each $k$. Carrying out the calculation on a computer, we find $6,10,15,21,28,36$ states for $k = 2,3,4,5,6,7$, respectively. This matches the formula $(k+1)(k+2)/2$ for the ground state degeneracy on a torus for the RR$k$ states~\cite{Read99}. 

\subsubsection{Filling fraction}

We can read off the filling fraction by considering the theory on a torus, which, as explained above, allows us to restrict the gauge fields to the Cartan subalgebra. From the equations of motion of the resulting Abelian Chern-Simons theory, we can obtain the Hall response, which determines the filling fraction to be
\begin{align}
\nu = q^T \tilde{K}^{-1} q = \frac{2}{k+2}.
\end{align}\\

\subsubsection{Shift}

We can also read off the shift by considering the coupling of the partons to the geometry of the space. Specifically, we couple the partons to an $SO(2)$ spin connection $\omega_\mu$, whose curl gives the curvature of the space. We note that noninteracting fermions residing in the LL indexed by $n$ carry an orbital spin $n+1/2$~\cite{Wen92}.

By considering the theory on a torus, which as above allows us to restrict the gauge fields to the Cartan subalgebra,
we see that integrating out the partons gives rise to an effective action
\begin{align}
  \mathcal{L} = &-\frac{k+1}{4\pi} \tilde{a} \partial \tilde{a} + \frac{1}{2\pi} \left(a + A + \frac{k+1}{2} \omega\right) \partial \tilde{a}
  \nonumber \\
 & - \frac{1}{4\pi} \text{Tr}\left[ \left(\mathcal{A}^I p^I + \frac{1}{2} \mathbb{I} \omega\right) \partial  \left(\mathcal{A}^I p^I + \frac{1}{2} \mathbb{I} \omega\right) \right]
  - \frac{1}{4\pi} \text{Tr}\left[ \left(\mathcal{A}^I p^I + \frac{3}{2} \mathbb{I} \omega\right) \partial \left(\mathcal{A}^I p^I + \frac{3}{2} \mathbb{I} \omega\right) \right].
\end{align}
Here $\mathbb{I}$ denotes the $k \times k$ identity matrix. In the first line of the above equation, the second CS term on the RHS has a coupling of $(k+1)/2$ between $\omega$ and $\tilde{a}$, which corresponds to the spin for the $1/(k+1)$ Laughlin state formed by the $b$ partons. The two terms in the second line of the above equation arise from the $n=0$ and $n=1$ Landau levels, respectively, that the $f$ fermions are filling. 
Simplifying, we obtain
\begin{align}
\label{eq:LagrangianNonSimple}\mathcal{L} =  -\frac{k+1}{4\pi} \tilde{a} \partial \tilde{a} + \frac{1}{2\pi} \left(a + A + \frac{k+1}{2} \omega\right) \partial \tilde{a}
  - \frac{2}{4\pi} \mathcal{A}^I \partial \mathcal{A}^J \text{Tr}\left[p^I p^J\right] - \frac{2}{2\pi}  \mathcal{A}^0 \partial \omega
    - \frac{10 k}{4} \frac{1}{4\pi} \omega \partial \omega ,
\end{align}
where we have used the fact that $\text{Tr } p^I = \delta_{I0}$. 
Recalling that $\mathcal{A}^0 = a$, Lagrangian (\ref{eq:LagrangianNonSimple}) can be written more compactly as
\begin{align}
  \mathcal{L} = &- \tilde{K}_{IJ} \tilde{\mathcal{A}}^I \partial \tilde{\mathcal{A}}^J + \frac{1}{2\pi} q^I A \partial \tilde{A}^I + \frac{1}{2\pi} s^I \omega \partial \tilde{A}^I
  \nonumber \\
&  - \frac{10 k}{4} \frac{1}{4\pi} \omega \partial \omega , \label{eq:LagrangianSimple}
\end{align}
where $\tilde{K}$, $\tilde{A}$, and $q$ are as defined above. The spin vector here is $s^T = ( (k+1)/2, -2,0,0,\dots, 0)$. 

Integrating out $\tilde{A}$ from Eq.~(\ref{eq:LagrangianSimple}) then gives
\begin{align}
\mathcal{L} = \frac{1}{4\pi} (q^T A + s^T \omega) \tilde{K}^{-1} \partial (q A + s \omega) - \frac{10 k}{4} \frac{1}{4\pi} \omega \partial \omega . 
\end{align}
The shift of the state on the sphere is defined to be~\cite{Wen92}:
\begin{align}
\mathcal{S} =2 \frac{q^T \tilde{K}^{-1} s}{q^T \tilde{K}^{-1} q} = 2 \nu^{-1} (\tilde{K}^{-1}_{11} \frac{k+1}{2} - 2 \tilde{K}^{-1}_{12}) .
\end{align}
One can verify that this gives
\begin{align}
\mathcal{S} = 1-k,
\end{align}
which agrees with the shift of the particle-hole conjugate of the Read-Rezayi states. 

We note that in order for the effective action above to have the complete coupling to the background geometry,
a gravitational Chern-Simons term proportional to the chiral central charge must also be added to capture the effect of the framing anomaly~\cite{Gromov15, Witten89}. However since this does not affect the shift, we have not included this in the above expressions. 

\subsection{Quasiparticle structure and fusion rules}

Consider a general $\nu = p/q$ FQH state, where $p$ and $q$ are coprime. On general grounds, the ground state degeneracy
must be a multiple of $q$. The ground state degeneracy on a torus equals the number of quasiparticles, $N_{\rm qp}$. 
The quasiparticles must break up into $N_{\rm qp}/q$ distinct sectors such that, within a given sector, the quasiparticles can be related to each other by inserting flux quanta. In other words, the quasiparticles of a FQH state at $\nu = p/q$ can always be decomposed as follows.
We have a set of $q$ topologically distinct quasiparticles described by
\begin{align}
V_a, \;\; Q_a = a p /q, \;\; \theta_a = \pi a^2 p/q, \;\; a = 0,\cdots, q-1.
\end{align}
These are Abelian quasiparticles with fractional charge $Q_a$ and exchange statistics $\theta_a$. 
The fusion rules of these Abelian quasiparticles are described by:
\begin{align}
V_a \times V_b = V_{a+b}.
\end{align}
Note that fusion with an electron yields a topologically equivalent quasiparticle, but changes the exchange statistics by $\pi$. 
{\it A priori}, we do not know if $V_q \simeq V_0$. 

The rest of the quasiparticles can be written as:
\begin{align}
\tau_\alpha V_a, \;\;\; \alpha = 0,\cdots, N_{\rm qp}/q - 1, \;\; a = 0,\cdots, q-1. 
\end{align}
In the case at hand, we have $\nu = 2/(k+2)$. Therefore we must treat the case where $k$ is even and $k$ is odd separately:
\begin{itemize}
 \item When $k$ is odd, the quasiparticles in our theory are of the form $\tau_\alpha V_a$, with $a = 0,\cdots, k+2$ and 
$\alpha = 0,\cdots, (k+1)/2 - 1$. 
 \item When $k$ is even, the quasiparticles are of the form $\tau_\alpha V_a$, with $a = 0,\cdots, (k+2)/2$ and 
$\alpha = 0,\cdots, k$. 
\end{itemize}
Since our effective field theory contains an $SU(k)_2$ CS term, the quasiparticle structure must inherit fusion rules associated with
$SU(k)_2$ CS theory. In particular, $SU(k)_2$ has the same fusion rules as $SU(2)_k$ by level-rank duality (see below for a brief review of $SU(2)_k$ fusion rules). 

\subsubsection{$k = 3$}
\begin{table}
\centering
\begin{tabular}{l | r |r r }
\hline
Label & Charge, $Q$ & Twist, $e^{i\theta}$ \\
\hline
$V^{{\bar 2}^3 1^{4}}_0$  & $0$ & $1$ &      \\
$V^{{\bar 2}^3 1^{4}}_1$  & $2/5$ & $e^{i\pi 2/5}$ &      \\
$V^{{\bar 2}^3 1^{4}}_2$  & $4/5$ & $-e^{i\pi 3/5}$ &      \\
$V^{{\bar 2}^3 1^{4}}_3$  & $1+1/5$ & $-e^{i\pi 3/5}$ &      \\
$V^{{\bar 2}^3 1^{4}}_4$  & $1+3/5$ & $e^{i\pi 2/5}$ &      \\
$V^{{\bar 2}^3 1^{4}}_5 \sim V^{{\bar 2}^3 1^{4}}_0 \Psi_e$ & $1$ & $-1$ \\
$V^{{\bar 2}^3 1^{4}}_6 \sim V^{{\bar 2}^3 1^{4}}_1 \Psi_e$ & $1 + 2/5$ & $-e^{i\pi 2/5}$ \\
$V^{{\bar 2}^3 1^{4}}_7 \sim V^{{\bar 2}^3 1^{4}}_2 \Psi_e$ & $1 + 4/5$ & $e^{i\pi 3/5}$ &      \\
$V^{{\bar 2}^3 1^{4}}_8 \sim V^{{\bar 2}^3 1^{4}}_3 \Psi_e$ & $1/5$ &  $e^{i\pi 3/5}$ &      \\
$V^{{\bar 2}^3 1^{4}}_9 \sim V^{{\bar 2}^3 1^{4}}_4 \Psi_e$ & $3/5$ & $-e^{i\pi 2/5}$ &      \\
\hline
\end{tabular}
\caption{Abelian particles of generic $\nu = 2/5$ state. Charge $2$ bosons are treated as completely trivial. Charge $1$ fermion (electron) is kept track of. We see that these properties match those of Table \ref{3antiRRtable} after a relabeling of particles. \label{k3partonParticles}}
\end{table}

The case $k = 3$ is particularly simple and can be treated explicitly. Here $N_{\rm qp} = 10$ and $q= 5$. We therefore have $10$ quasiparticles, described by:
\begin{align}
&V^{{\bar 2}^3 1^{4}}_a , \;\;Q^{{\bar 2}^3 1^{4}}_a = 2a/5, \;\; \theta^{{\bar 2}^3 1^{4}}_a = 2\pi a^2/5,~a = 0,\cdots 4
\nonumber \\
&\tau V^{{\bar 2}^3 1^{4}}_a , \;\; Q^{{\bar 2}^3 1^{4}}_a = 2a/5, \;\; a = 0,\cdots, 4. 
\end{align}

Since the chiral central charge is not an integer, the theory must be non-Abelian. Therefore $\tau$ must be a non-Abelian particle.
The non-Abelian part of the fusion rules must be associated with the fusion rules of $SU(2)_3$; the non-Abelian fusion rules of
$SU(2)_3$ are derived from the Fibonacci theory \cite{DiFrancesco97,Rowell09}. 

In other words, any non-Abelian particle must be of the form $\tau = \sigma X$, where $\sigma \times \sigma = 1 + \sigma$ 
is a non-Abelian Fibonacci particle, and $X$ is the Abelian part which encodes any additional $U(1)$ phases in the statistics. This means that 
$\tau \times \tau = X^2 + X^2 \sigma = X^2 + X \tau$. Since $\tau$ was already a particle in the theory, this means that $X$ must 
also correspond to some Abelian particle of the theory. Therefore $X$ must correspond to one of the fields $V_a$. This, in turn,
implies that $\sigma$ itself must be a particle of the theory! For simplicity, we relabel $\sigma \rightarrow \tau$. 

Thus we see that $\tau \times \tau = 1+ \tau$. Therefore we have the complete set of fusion rules. 
The above implies that our theory has a subcategory which is a braided fusion category with fusion rules $\tau \times \tau = 1 + \tau$. By consistency of the braided fusion category (i.e., by using the ribbon identity \cite{Wang10}), the only possible choices for $\theta_\tau$ are $\theta_{\tau}  =\pm 2/5 (2\pi)$. Furthermore, the only electric charge for $\tau$ that is consistent with its fusion rules are $Q_\tau = 0$. 

We would like to also determine the statistics of $\tau V_a$. Let $\theta_{\tau, 1}$ be the mutual statistics between $\tau$ and $V_1$. 
It follows from the ribbon identity that 
\begin{align}
\theta_{\tau V_a} = \theta_{\tau} + \theta_{V_a} + a \theta_{\tau, 1} . 
\end{align}
Since $V_1$ is just a quasiparticle obtained by inserting a single flux quantum, while $\tau$ is an electrically neutral non-Abelian quasiparticle, we should have $\theta_{1,\tau}  = 0$. 

By comparing with Table \ref{3antiRRtable} and the discussion in Sec.~\ref{3aRRsec}, we can see that these properties of the quasiparticles match exactly those of the $k=3$ anti-Read Rezayi state. 

Thus we only have one ambiguity left, which is the sign of $\theta_\tau$. In principle we can directly compute this using the edge theory. However only one choice of $\theta_\tau$ is consistent with the chiral central charge $c_{-} = -4/5 = 2 - 14/5$. We can see this as follows. First, we observe that our theory decomposes as $\mathcal{A}_1 \times \mathcal{A}_2$, where $\mathcal{A}_1$ is a non-Abelian theory with two types of particles, $\{1,\tau\}$ and $\tau \times \tau = 1 + \tau$. $\mathcal{A}_2$ is an Abelian theory whose properties are summarized in Table~\ref{k3partonParticles}. Note in particular that $\mathcal{A}_2$ is the topological order of the usual $\nu = 2/5$ Abelian hierarchy state. Therefore $\mathcal{A}_2$ is by itself a (spin) modular tensor category. Since our theory effectively splits into two independent theories, the chiral central charge is $c_{-} = c_{-,1} + c_{-,2}$, with $c_{-,2} = 2$. Therefore we need $c_{-,1} = -14/5$ in order to match the central charge $c_{-} = -4/5$ derived from the calculation in the edge theory below.  

For $c_{-,1} = -14/5$, the topological spin of the Fibonacci particle is uniquely specified to be $\theta_\tau = -2/5$~\cite{Rowell09}. Note that the above considerations fully specify the modular $S$ matrix of the theory in addition to the modular $T$ matrix.

\subsubsection{Review of $SU(2)_k$ fusion rules}
Recall that the particles of $SU(2)_k$ can be labeled as $[j/2]$, which correspond to the spin $j/2$ representation of $SU(2)$,
for $j = 0,\cdots, k$. These have fusion rules
\begin{align}
[j/2] \times [j'/2] = \sum_{l = |j-j'|}^{{\rm min}((j+j')/2, (k-(j+j'))/2)} [l] .
\end{align}
In particular, 
\begin{align}
[k/2] \times [k/2] = [0].
\end{align}
Thus $[k/2]$ is an Abelian particle. In general, we have
\begin{align}
[j/2] \times [j/2] = [0] + [1] + \cdots + [{\rm min}(j, k-j)] .
\end{align}
Therefore all particles except for $[k/2]$ are non-Abelian. 

Since $[k/2]$ has $\mathbb{Z}_2$ fusion rules, this also means that there must be $(k+1)/2$ particles that cannot be obtained from each other by fusing with the Abelian $[k/2]$ particle. 

As an example, $SU(2)_3$ has $4$ particles, with the following fusion rules:
\begin{align}
1/2 \times 1/2 &= 0 + 1
\nonumber \\
1 \times  1 &= 0 + 1
\nonumber \\
3/2 \times 3/2 &= 0
\nonumber \\
1 \times 1/2 &= 1/2 + 3/2.
\end{align}

\subsection{Edge theory}

The mean-field ansatz of the partons is described by $k+1$ free chiral (say, right-moving) fermions and $2k$ chiral left-moving fermions. This can be described by a $U(k+1)_1 \times U(2k)_{-1}$ Wess-Zumino-Witten (WZW) conformal field theory (CFT). The gauge projection consists of projecting to the $SU(k+1)$ invariant sector for the first $k+1$ fermions, the $SU(k)$ invariant sector for the left-moving fermions, together with a $U(1)$ projection.
In other words the CFT is a $\frac{U(k+1)_1 \times U(2k)_{-1}}{SU(k+1)_1 \times SU(k)_{-2} \times U(1)_{-1}}$ coset theory. 
Alternatively, since $U(k)_1/SU(k)_1 \simeq U(1)_k$, this is equivalent to a $\frac{U(1)_{k+1} \times U(2k)_{-1}}{SU(k)_{-2} \times U(1)_{-1}}$
coset theory.

\subsubsection{Central charge}
The total chiral central charge can be read off from the above coset theory as
\begin{align}
c_{-} = c_{-,{\rm MF}} - c_{-,{\rm gauge}} .
\end{align}
Here $c_{-,{\rm MF}}$ is the chiral central charge of the mean-field (MF) ansatz of the partons, 
\begin{align}
c_{-,{\rm MF}} = (-2k)+(k+1) = -k+1 .
\end{align}
Since the gauge effective action is an $SU(k+1)_1 \times SU(k)_{-2} \times U(1)_{-1}$ CS theory, we can read off $c_{-,{\rm gauge}}$:
\begin{align}
c_{-,{\rm gauge}} = k -\frac{2(k^2-1)}{k+2} - 1.
\end{align}
Therefore,
\begin{align}
c_{-} = -k +1 -k  + \frac{2(k^2-1)}{k+2} + 1 = 1 - \frac{3k}{k+2},
\end{align}
which precisely matches the total chiral central charge of the particle-hole conjugate of the $k$-cluster RR state. \\

In conclusion, we have shown that the filling factor, shift on the sphere, chiral central charge, ground state degeneracy on a torus and the non-Abelian part of the fusion rules (for $k\geq 2$) for the quasiparticles of the parton states given in Eq.~(\ref{wfnAnsatz}) are the same as those of the aRR$k$ states. For $k =3$, we further demonstrated that the full anyon content, including all the fusion rules and the topological spins, is identical to those of the aRR$3$ state. These results strongly suggest that the our parton states given in Eq.~(\ref{wfnAnsatz}) and the aRR$k$ states describe the same topological phases.

\section{Comparison of the ``$\bar{n}\bar{2}^{2}1^{4}$'' parton sequence with other families of candidate states occurring at the same sequence of filling factors}
\label{sec:comparison_parton_others}
In the main text we proposed a new ``$\bar{n}\bar{2}^{2}1^{4}$'' parton sequence to capture some of the observed states in the SLL. The ``$\bar{n}\bar{2}^{2}1^{4}$'' parton states are described by the wave functions:
\begin{equation}
\Psi^{\bar{n}\bar{2}^{2}1^{4}}_{n/(3n-1)} = \mathcal{P}_{\rm LLL} [\Phi^{*}_{n}]  [\Phi^{*}_{2}]^{2} \Phi^{4}_{1}\sim \frac{\Psi^{\rm CF}_{n/(2n-1)}[\Psi^{\rm CF}_{2/3}]^{2}}{\Phi^{2}_{1}},
\label{eq:parton_SM}
\end{equation}
which in the spherical geometry occur at a shift $\mathcal{S}^{\bar{n}\bar{2}^{2}1^{4}}=-n$. The $n/(3n-1)$ sequence of filling factors matches that of the Bonderson-Slingerland (BS) states~\cite{Bonderson08}, which are built up from the Pfaffian state. The shift of the $n/(3n-1)$ BS state, $\mathcal{S}^{\rm BS}=4-n$, is different from that of the parton state given in Eq.~(\ref{eq:parton_SM}). Therefore the parton and BS states have different Hall viscosities~\cite{Read09}, $\eta_{\rm H} = \hbar \nu \mathcal{S}/(8\pi\ell^{2})$. Furthermore, these states feature different thermal Hall conductances. In particular, the $2/5$ BS state has~\cite{Bishara08} $\kappa_{xy}=(1/2)[\pi^2 k_{\rm B}^2 /(3h)]T$, while the $\bar{2}^{3}1^{4}$ parton state has $\kappa_{xy}=-4/5[\pi^2 k_{\rm B}^2 /(3h)]T$ (the two lowest filled LLs with spin up and spin down provide an additional contribution of $2[\pi^2 k_{\rm B}^2 /(3h)]T$ to $\kappa_{xy}$). Thus the parton states of Eq.~(\ref{eq:parton_SM}) are topologically distinct from the BS states.\\

The anti-Pfaffian analog of the BS state has a shift of $\mathcal{S}^{\rm aPf-BS}=-n$, which is the same as that of the parton state given in Eq.~(\ref{eq:parton_SM}). This suggests that the two states could lie in the same phase. However, the thermal Hall conductance of the anti-Pfaffian analog of the $2/5$ BS state is $-3/2[\pi^2 k_{\rm B}^2 /(3h)]T$~\cite{Bishara08}, which differs from that of the parton state. Thus these states possess different topological order (and are experimentally distinguishable), despite having the same shifts and filling factors.\\

We mention here that Jolicoeur~\cite{Jolicoeur07} has also proposed states along the sequence $n/(3n-1)$. These states occur at shift $\mathcal{S}^{\rm Jol}=1$ and are therefore topologically different from our parton states. The Jolicoeur wave functions at $2/5$ and $3/8$ arise from RR states involving four and six clusters, respectively. These wave functions are not easily amenable to a numerical calculation. To the best of our knowledge their properties have not been studied in detail in the literature.

\end{document}